\documentclass[aps,pra, reprint,notitlepage,floatfix]{revtex4-2}
\usepackage{placeins} 

\usepackage{graphicx}
\usepackage{dcolumn}
\usepackage{bm}
\usepackage{braket} 

\usepackage{amssymb}
\usepackage{amsmath}
\usepackage{multirow}

\usepackage{xcolor, soul}
\sethlcolor{yellow} 

\usepackage{tikz}
\usetikzlibrary{datavisualization}
\usepackage{pgfplots}
\usepackage{pgfplotstable}
\usepgfplotslibrary{colormaps}
\pgfplotsset{compat=1.18}

\usepackage{lipsum}

\makeatletter
\def\thanks#1{\protected@xdef\@thanks{\@thanks
        \protect\footnotetext{#1}}}
\makeatother
\begin{document}
\usepgfplotslibrary{groupplots}
\usepgfplotslibrary{fillbetween}

\title{Fiber-coupled broadband quantum memory for polarization-encoded photonic qubits}

\author{Sandra Cheng}
\email{scheng5@umbc.edu}
\author{Carson Evans}

\author{Todd Pittman}

\affiliation{Department of Physics and Quantum Science Institute, University of Maryland Baltimore County, Baltimore, MD 21250, USA}

\date{\today}

\begin{abstract}
Various near-term quantum networking applications will benefit from low-loss, fiber-coupled photonic quantum memory devices with high efficiencies. We demonstrate a fiber-coupled loop-and-switch quantum memory platform with a pass-through efficiency of $\sim$54\% and an overall storage efficiency that scales as $\sim\! 0.5^{N+1}$ where $N$ is the number of storage cycles.  We highlight the trade-off between memory lifetime and qubit accessibility in this platform by using two different storage cycle times of $\sim\!$ 40~ns and $\sim\! 0.5~\mu$s, and demonstrate high-fidelity storage and retrieval of ultra-broadband single-photon polarization qubits in both cases.
\end{abstract}

\maketitle

\section{\label{sec:intro}Introduction}

The past two decades have seen tremendous progress in the realization of photonic quantum memory devices~\cite{lvovsky_optical_2009, Simon_2010, heshami_quantum_2016, SHINBROUGH2023297}, with a recent emphasis on intrinsically fiber-coupled devices suitable for ``plug-and-play'' quantum networks~\cite{kimble_qInternet_2008,Qnetworksreview2022}. While \textit{emissive} memories can exhibit near-unit fiber coupling efficiencies~\cite{higheffcqedtrappedion1, higheffcqedtrappedion2}, the situation can be more challenging for \textit{absorptive} memories where photons arriving from a distant source must be stored and then later released. Nonetheless, fiber-coupled absorptive memories have recently been demonstrated using the atomic ensemble platform~\cite{higheffatomicEnsemble1}, solid-state platform~\cite{higheffSolidState1, higheffSolidState2} and fiber cavity platform~\cite{fiberCavity2024}. Consequently, there is currently a need for investigation into other absorptive memory platforms that may also be useful for fiber-coupled applications.

The loop-and-switch (LAS) absorptive memory platform~\cite{cqm_2002, 2013-furusawa, freeSpaceLoopQM_5_makino_2016, kaneda_quantum-memory-assisted_2017, freeSpaceLoopQM_1_2019, freeSpace_Kaneda_SPgen_2019, 2020-cqm-and-atomic,freeSpaceLoopQM_7_meyer-scott_SPgen_2022, freeSpaceLoopQM_8_2022, guo2023multiplex, 2023npj_CQMentanglement, 2023Evans, 2024-silberhorn,opticalFibersQM_4_Fook_Lee_2024, hanamura2025scalableopticalquantumstate, sun2025freeSpaceLoop8} is particularly appealing due to its reliance on underlying technologies which can be largely modular, off-the-shelf, and fiber-coupled. However, some of these fiber-coupled components --- particularly the vital voltage-controlled switches --- tend to be lossy and/or slow compared to their free-space counterparts, resulting in limitations on the application spaces for these memories. For example, Lee \textit{et al.}~\cite{opticalFibersQM_4_Fook_Lee_2024} recently demonstrated a fully fiber-based LAS memory with a remarkably high storage efficiency of $\sim$$0.8^{N+1}$, where $N$ is the number of loop cycles. However, this memory utilized a commercially available fiber-coupled switch with a maximum repetition rate of only $\approx$~100~kHz, thereby restricting the minimum storage cycle time to comparatively long values of $\sim$10$~\mu$s. While this enabled high-fidelity ultralong storage times, much shorter cycle times and higher repetition-rate switches ($\sim$MHz to GHz) are desirable for various other types of ultrafast quantum memory network applications~\cite{pittman_heralded_2003, freeSpace_Kaneda_SPgen_2019, freeSpaceLoopQM_7_meyer-scott_SPgen_2022, guo2023multiplex}.  Ideally, the fiber-coupled LAS platform would boast a short rise-time, low-loss, and high-rate switch that could be seamlessly used in both the ultralong and ultrafast fiber-loop storage regimes.

Here we report on such a platform by using a hybrid approach that involves a high-speed free-space electrically controlled switch (rise time $\sim$10~ns, repetition rate $\sim$1~MHz) with a single mode fiber (SMF) storage line. Our platform also uses a low-loss free-space optical circulator with input and output channels coupled to SMF. The end result is a fully fiber-coupled plug-and-play LAS memory that can be easily reconfigured for applications in a wide range of timescales.  We observe a pass-through transmission of $(54.1 \pm 0.2)\%$ and a storage efficiency that scales as $\sim$$0.5^{N+1}$. In addition, we perform quantum state tomography to confirm the high-fidelity storage and retrieval of polarization-encoded single-photon qubits for active storage times ranging from $\sim$40~ns to 1.5~$\mu$s.

\begin{figure*}[hbt]
    \centering
    \includegraphics[width=0.6\linewidth]{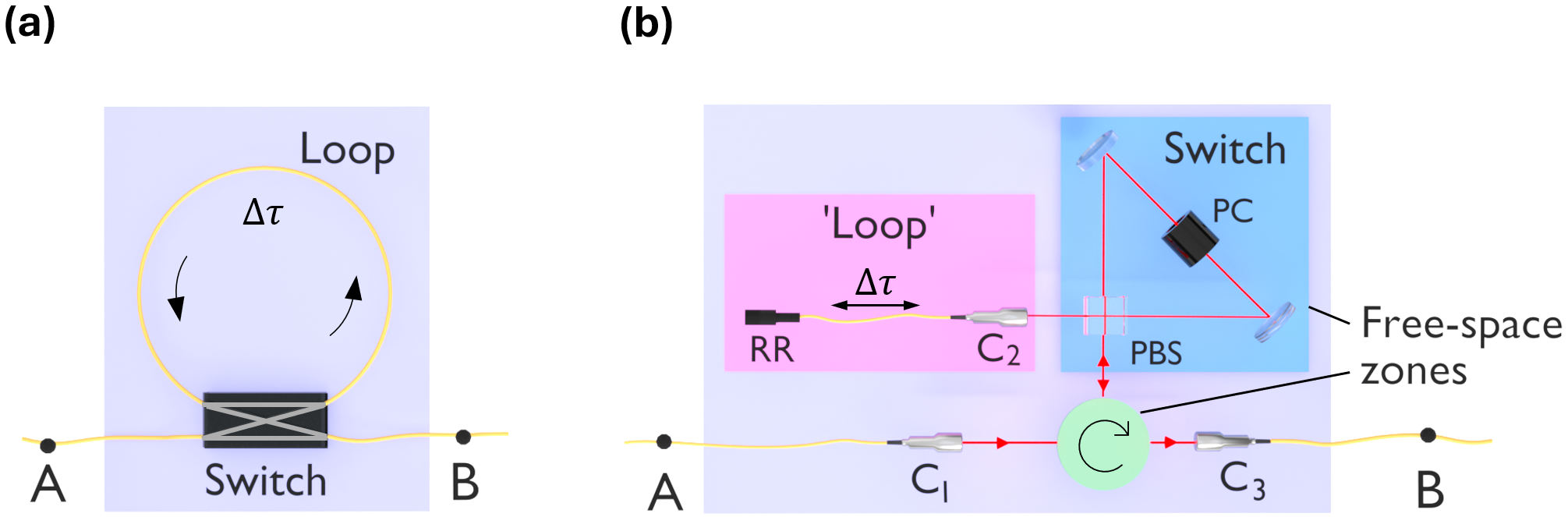}
    \caption{(a) Conceptual overview of the fiber-coupled loop-and-switch (LAS) quantum memory platform. Incident photonic qubits are stored for a total time of $N \Delta \tau$, where $N$ is the user-controlled number of round trips. (b) Overview of our hybrid implementation of a fiber-coupled LAS memory which uses both free-space (green and blue zones) and fiber (pink zone) components. The storage loop is a single-mode fiber terminated by a retroreflector (RR), while the polarization-insensitive switch is comprised of a Pockels cell (PC) and a polarizing beamsplitter (PBS) set in a Sagnac interferometer. The switch and circulator are both free-space units. The key fiber-to-free-space couplers are labeled $C_1$, $C_2$ and $C_3$.}
    \label{fig:loop-and-switchDiag}
\end{figure*}

Figure~\ref{fig:loop-and-switchDiag}(a) shows a conceptual overview of the LAS quantum memory platform in its simplest form. While similar to the well-known addressable fiber-loop classical optical memory~\cite{Avramopoulos:93, langenhorst1996opticalfiber}, our LAS quantum memory features a crucial distinction --- a carefully designed switch that preserves the state of arbitrary polarization-encoded qubits~\footnote{Because time-bin qubits can, in principle, be used with conventional optical switches~\cite{sun2025freeSpaceLoop8, kaneda_quantum-memory-assisted_2017}, an alternative approach involves polarization to time-bin qubit transduction~\cite{timebintransduction-1999, Kupchak_timebintransduction2017} combined with conventional switching~\cite{freeSpaceLoopQM_8_2022}}. In the ideal case, an incident photon is switched into the storage loop, where it is safely stored until the user actively releases it after $N$ round trips. For clarity, here we define the end-to-end efficiency $\eta_N$ as follows: given a photon at point A in the input fiber, $\eta_N$ is the probability that the photon will be found at point B in the output fiber after being stored for $N$ cycles. $\eta_0$ is thus the pass-through transmission when the switch remains off. Due to absorption and scattering losses in the switch and storage loop, it is clear from Fig.~\ref{fig:loop-and-switchDiag}(a) that $\eta_N$ will degrade as a function of $N$. 

Fig.~\ref{fig:loop-and-switchDiag}(a) also highlights a fundamental trade-off between storage time and qubit accessibility in the LAS platform, which can be optimized for various timescale applications by simply adjusting the round trip loop time $\Delta \tau$.  For protocols emphasizing longer storage times~\cite{Muralidharan_2016,opticalFibersQM_4_Fook_Lee_2024}, large $\Delta \tau$ values and slower switching speeds can be used, while high-accessibility applications~\cite{freeSpace_Kaneda_SPgen_2019, freeSpaceLoopQM_7_meyer-scott_SPgen_2022, guo2023multiplex, hanamura2025scalableopticalquantumstate} require the shortest possible $\Delta \tau$ values and accordingly, the fastest switching speeds. For perspective, probabilistic operations of the kind in Refs.~\cite{freeSpace_Kaneda_SPgen_2019, freeSpaceLoopQM_7_meyer-scott_SPgen_2022, guo2023multiplex, hanamura2025scalableopticalquantumstate} typically involve higher-order interference effects among multiple photon sources driven by a common $\sim$100~MHz clock cycle (derived from standard mode-locked lasers). Therefore, in order to enhance these operations using fiber-coupled LAS memory networks, switching speeds and minimum loop cycle times on the order of $\sim$10~ns are beneficial.  In principle, as long as the switching speed is shorter than the minimum required loop storage time, the loop time itself can then be easily increased to match the clock cycle associated with any slower applications. In this work, we demonstrate this idea by using a fast switching speed of $\sim$10~ns with two different loop times of $\Delta \tau_s$~=~36.5 ns (short) and $\Delta \tau_l = 0.526~\mu$s (long) in an easily reconfigurable architecture.

Figure~\ref{fig:loop-and-switchDiag}(b) shows an overview of our implementation of the required polarization qubit-preserving switch using the original design of Pittman and Franson~\cite{cqm_2002}. Here, a single photon qubit of the form $\ket{\psi} = \alpha \ket{H} + \beta \ket{V}$ is injected into a polarizing beamsplitter (PBS)-based Sagnac interferometer and delocalized into counter-propagating horizontal and vertical components which coherently recombine upon exit. Consequently, active storage and release is controlled by the user through carefully timed bit flips (i.e. $\ket{H} \leftrightarrow \ket{V}$) applied within the interferometer using a high-speed Pockels cell (PC) biased to its half-wave voltage.     

In analogy with Fig.~\ref{fig:loop-and-switchDiag}(a), the Sagnac interferometer in Fig.~\ref{fig:loop-and-switchDiag}(b) essentially forms the switch, while the storage `loop' is a simple out-and-back fiber terminated by a retroreflector (RR). Note that due to the weakly dispersive nature of the switching mechanism --- the Pockels effect --- this fiber-coupled LAS memory is inherently broadband, enabling us to store heralded single-photons with $\sim$10~THz bandwidths produced by a spontaneous parametric down-conversion (SPDC) source. In this work we utilize photons centered at 780~nm, but the non-resonant nature of the switching mechanism means this overall memory architecture can be implemented for telecom-band photons as well. This feature of the overall LAS platform stands in contrast to matter-based quantum memories which rely on specific atomic or solid-state resonant frequencies~\cite{Simon_2010, heshami_quantum_2016}.

\section{\label{sec:exp}Experiment} 
As illustrated in Figure~\ref{fig:loop-and-switchDiag}(b), this fiber-coupled LAS memory is a hybrid device; the input, output and storage channels are SMF, while the switch and circulator are formed by free-space bulk-optic interferometers that enable higher-speed and lower-loss operation compared to typical commercially available 780~nm fiber-based devices. From a practical experimental point of view, the overall alignment of this hybrid optical setup is facilitated by approximating the SMF modes launched from three key fiber couplers $C_1, C_2$, and $C_3$ as Gaussian intensity distributions in free space, and then mode-matching these three beams by ensuring their beam waists are centered in the PC. This enables simultaneous optimization of both the $C_1$-to-$C_3$ coupling needed for high pass-through efficiency $\eta_0$, as well as the $C_1$-to-$C_2$, repeated $C_2$-to-$C_2$, and $C_2$-to-$C_3$ couplings needed to store qubits for $N$ round trips with high end-to-end efficiency $\eta_N$.

\begin{figure*}[th] 
\centering
\includegraphics[width=\textwidth]{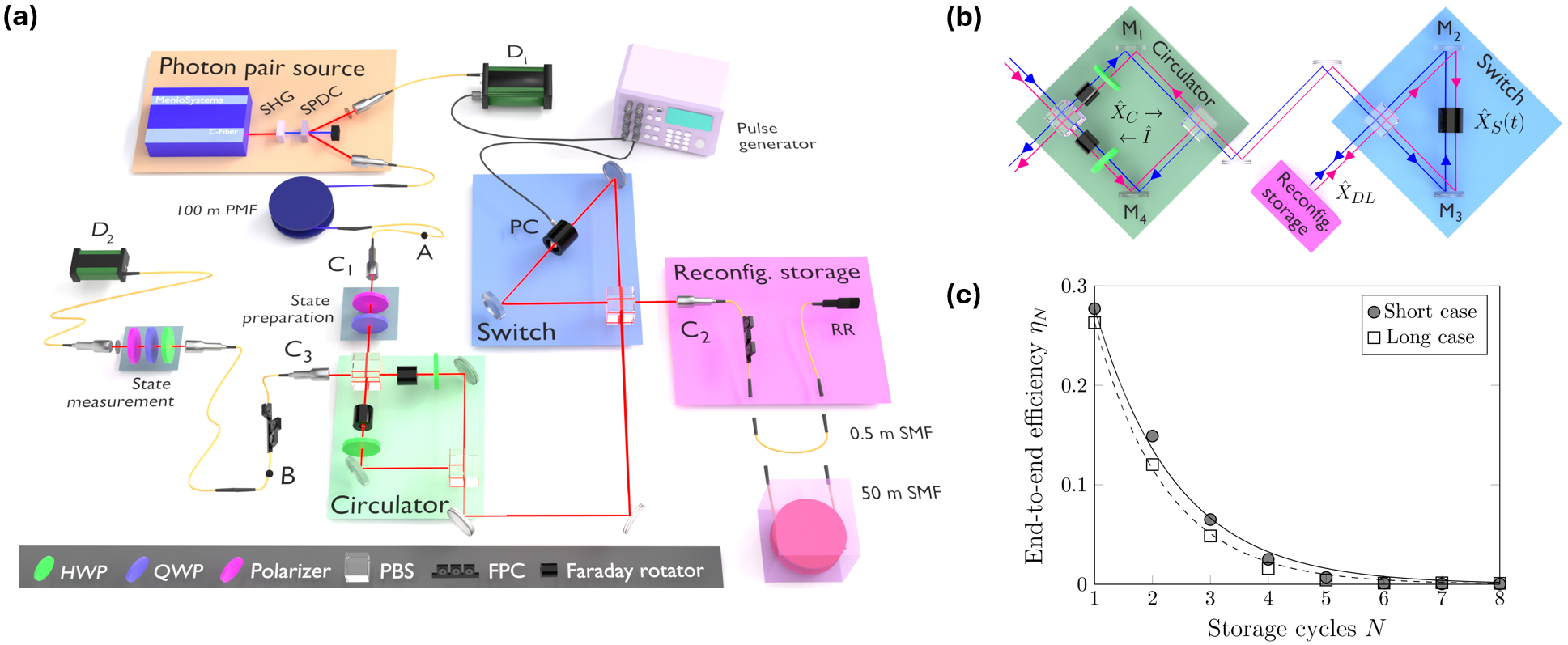} 
\caption{(a) Experimental schematic: Heralded single photons from a type-I SPDC source are delivered via single-mode fiber (SMF) into the hybrid LAS memory. The circulator (green free-space zone) consists of Faraday rotators in a polarizing beamsplitter (PBS)-based Mach-Zehnder interferometer while the switch (blue free-space zone) consists of a Pockels cell (PC) in a PBS-Sagnac interferometer. A reconfigurable SMF storage line (pink fiber zone) with two different fiber lengths (short: 0.5~m, long: 50~m) enables a demonstration of flexible memory operation on two different timescales. Heralding of the SPDC idler initiates active user-controlled storage and release driven by a pulse generator. The free-space circulator routes released photonic polarization qubits into an SMF output channel for fully fiber-coupled memory operation. (b) Conceptual overview of the $\ket{H}$ (red arrows) and $\ket{V}$ (blue arrows) paths through the `figure eight' (F8) common-path interferometer, and the bit flip operations ($\hat{X}_C, \hat{X}_S(t), \hat{X}_{DL}$) associated with each of the three zones. (c) Experimental measurements showing memory efficiencies of roughly $\sim\!50\%$ ($\sim\!44\%$) per cycle for the short (long) storage cases using the $\ket{H}$ input state, with solid (dashed) lines representing fits using Eq.~\ref{eq:eff}. Additional details are included in the main text.}
\label{fig:expSetup-f8-eff}
\end{figure*}

The detailed experimental setup is depicted in Figure~\ref{fig:expSetup-f8-eff}(a). A 100 MHz repetition rate pulsed laser (Menlo Systems C-Fiber 780; central wavelength 780~nm, pulse width $\sim$100~fs) pumps a beta barium borate (BBO) crystal to create 390~nm photons via second-harmonic generation (SHG). These blue photons pump another BBO crystal producing degenerate photon pairs centered at 780~nm via type-I SPDC. Both downconverted photons are coupled into SMFs and the idler is first detected by a single-photon avalanche diode $D_1$, producing an electronic signal which triggers a pulse generator (BNC Model 575) used to control the PC (QUBIG GmbH Model HVOS-NIR; rise time $\approx$~10~ns). 
To compensate for the cumulative electronic latency of $\approx$~240~ns in the $D_{1}$-to-PC channel, the signal photon first passes from its initial 1~m SMF through two 50~m polarization-maintaining fiber (PMF) patchcords (connected with their fast and slow axes perpendicular to each other) with a propagation time of $\approx$~495~ns. The signal then passes through an additional 1~m SMF and launches out of fiber coupler $C_1$ into the quantum memory. The use of multiple plug-and-play fiber patchcords in the signal photon delay channel facilitates the injection of auxiliary fiber-coupled light sources into the memory for various alignment and loss measurement purposes.

A state preparation stage consisting of a rotatable polarizer and quarter wave-plate (QWP) following $C_1$ is used to negate the effects of birefringence in the signal photon delay channel, and prepare the standard set of three single-photon polarization states $\{ \ket{H}, \ket{D} = \frac{1}{\sqrt{2}}(\ket{H}+\ket{V})$, $\ket{R} = \frac{1}{\sqrt{2}}(\ket{H}-i\ket{V})\}$ used to test the ability of the memory to store arbitrary qubits. These qubits then enter the free-space circulator, formed by a PBS-based Mach-Zehnder interferometer (PBS-MZ) with a Faraday rotator (Thorlabs I780R5) and half wave-plate (HWP) in each arm. Note that the circulator is set to implement the forward-direction non-reciprocal bit flip (denoted $\hat{X}_{C}$) needed to route the qubits into the switch. After storage of $N$ cycles, the reverse direction implements an identity operation used to route the released qubits into the fiber-coupled memory output port $C_{3}$. A state measurement stage consisting of a free-space U-bench equipped with additional waveplates and a polarizer, followed by single-photon avalanche diode $D_{2}$, is used to perform quantum state tomography on the stored qubits~\cite{measofQubits}.  Both $D_{1}$ and $D_{2}$ (Excelitas SPCM-AQ4C) are equipped with 25 nm ($\sim$10~THz) bandwidth interference filters (IFs) centered near 780~nm. Coincidence counting between the two detectors was performed using a high-resolution timetagger (IDQuantique 900) with a coincidence window of 4~ns.\\ \indent
The overall storage protocol for $N$ cycles involves the user-controlled dynamic switching bit flip operations (denoted $\hat{X}_{S}(t)$) implemented by the PC, and an additional bit flip operation (denoted $\hat{X}_{DL}$) implemented by a fiber polarization controller (FPC) which controls the total birefringence of the entire reconfigurable fiber delay line. The delay line is comprised of a 1~m SMF patchcord, a 1~m fiber RR (Thorlabs P1-780R-P01-1), and the choice of a 0.5~m or 50~m SMF patchcord to produce the short and long storage cycle times of $\Delta \tau_s$~=~36.5~ns and $\Delta \tau_l$~=~0.526~$\mu$s respectively. With the latter, we observed time-dependent birefringence due to thermal and mechanical fluctuations for storage times $\gtrsim$ 550~ns in our lab. To mitigate these effects, the 50~m spool was placed in an insulated box during data collection. \\ \indent
Although the delay line bit flip operation $\hat{X}_{DL}$ is not required for basic operation, it enables a significant experimental benefit: the combined effects of $\hat{X}_{DL}$, $\hat{X}_{C}$, and $\hat{X}_{S}(t)$ essentially convert the combined PBS-MZ and PBS-Sagnac into one large overall phase-stable common-path interferometer that we refer to as a `figure eight' (F8) interferometer. Figure~\ref{fig:expSetup-f8-eff}(b) provides a conceptual overview of the F8 operation. Referring to the four labeled mirrors, it can be seen that for any qubit $\ket{\psi} = \alpha \ket{H} + \beta \ket{V}$ stored for $N$ cycles, the overall paths taken by its $\ket{H}$ and $\ket{V}$ components are, respectively, described by:
\begin{alignat*}{1}
    (M_4) \rightarrow (M_2 \rightarrow M_3 \rightarrow \textrm{storage})^N \rightarrow (M_2 \rightarrow M_3 \rightarrow M_1) \\
    (M_1) \rightarrow (M_3 \rightarrow M_2 \rightarrow \textrm{storage})^N \rightarrow (M_3 \rightarrow M_2 \rightarrow M_4)
\end{alignat*}

\noindent Because each initial component traverses the same optical path but in reverse directions, the PBS-MZ arms need not be balanced within the coherence length of the single photon qubit. Additionally, any PBS-MZ path-length drifts occurring during experimental data-collection result in inconsequential overall phase shifts, rather than deleterious phase-shift errors on the stored qubits. This latter technique is valid as long as the PBS-MZ phase drifts are slower than the longest desired memory storage time, which is the case in our experiments; we observe typical phase drifts of $2\pi$ on the order of several minutes, while our longest storage times are on the order of several microseconds.

The timing protocol for the $\hat{X}_{S}(t)$ application is as follows: for the baseline $N=0$ ($N=1$) storage cases, the PC is simply turned on (off) the entire time. For the more general case of $N\geq2$, the PC is turned on immediately after the first passage through the switch, and then turned off just before the $N^{\textrm{th}}$ passage. Note that for all $N$ round trips, the combination of the repeated $\hat{X}_{S}(t)$ and $\hat{X}_{DL}$ bit-flipping leads to an automatic cancellation of phase shift errors that build up from time-independent, static birefringence within the memory. In addition, bit flip errors (e.g. polarization rotations $\gtrless 90^\circ$ by the PC) result in early ejection of qubits from the memory, and manifest as loss.

As will be seen in the next section, the combination of (1) the common-path F8 phase drift removals, (2) the birefringent phase shift error cancellations, and (3) the bit flip error filtering effects enable the observation of polarization-based qubit fidelities that do not significantly degrade as a function of storage time. Consequently, the primary limiting factor in this fiber-coupled LAS memory is simply the overall loss per storage cycle.

\section{\label{sec:results}Results} 
Figure~\ref{fig:expSetup-f8-eff}(c) summarizes this overall loss per cycle by plotting the experimentally measured end-to-end memory efficiency $\eta_{N}$ for $N \in [1,8]$ storage cycles. To facilitate these measurements, we decompose the overall storage path into segments separated by the three fiber couplers $C_{i}$, and characterize these segments by experimentally accessible transmission parameters $\gamma_{i:j}$, with $i,j = (1,2,3)$. These parameters include the $C_{i}$-to-$C_{j}$ free-space to fiber coupling losses as well as the loss due to any components between $C_{i}$ and $C_{j}$. 

For example, $\gamma_{2:3}$ includes the losses in the two free-space zones, which includes minor absorptive losses associated with the multiple PBSs and mirrors ($\approx 7\%$ combined) as well as the more significant transmission loss through the PC ($\approx 10\%$), plus ejection loss in the PBS-MZ due to imperfect polarization rotations by the combined Faraday rotators and half-waveplate units (total $\approx 15\%$ including transmission losses). In a similar way, $\gamma_{1:2}$ is defined to include the losses from both free-space zones as well as the out-and-back loss in the fiber delay zone, while $\gamma_{2:2}$ is defined to include only the free-space switch zone and the fiber zone losses. This fiber zone loss consists of the fiber RR loss ($\approx19\%)$, fiber connector losses (total $\approx15\%$), and fiber attenuation losses associated with the fiber delay lines ($\approx$~0.01\% and $\approx$~8.8\% loss for the short and long lengths respectively). In order to best simulate well-prepared qubits arriving at the memory input during realistic quantum networking operation, we intentionally neglect the losses caused by the state preparation stage of Figure~\ref{fig:expSetup-f8-eff}(b) in our values of $\gamma_{1:2}$ and $\gamma_{1:3}$.

With these transmission parameters, $\eta_N$ can be neatly described as:
\begin{equation}
\eta_N =
\begin{cases}
  \gamma_{1:3} & \text{$N$ = 0}    \\
  \gamma_{1:2} \times \gamma_{2:2}^{N-1} \times \gamma_{2:3} & \text{$N$ $\geq$ 1} \label{eq:eff}\\
\end{cases}
\end{equation}

Using a power meter and an auxiliary 780~nm pulsed laser that has a spectral bandwidth comparable to that of the single-photon qubits, we measured $\gamma_{1:3} = 0.541 \pm 0.002$, $\gamma^S_{1:2} = 0.419 \pm 0.007$, $\gamma^L_{1:2} = 0.398 \pm 0.005$, and $\gamma_{2:3} = 0.662 \pm 0.007$. These values were found to be the same to within experimental error for all polarization states.

The $\gamma_{2:2}$ parameter values were extracted from SPDC coincidence counting rate data using a variety of polarization states for $N\in[1,8]$. Figure~\ref{fig:expSetup-f8-eff}(c) shows example data for $\ket{H}$ input states. Exponential fits to the raw coincidence counting decay rates provided  $\gamma_{2:2}^{H} = 0.49 \pm 0.03$ for the short storage case, and $\gamma_{2:2}^{H} = 0.43 \pm 0.02$ for the long storage case. Analogous data for $|D\rangle$ input states gave $\gamma_{2:2}^{D} = 0.50 \pm 0.03$ for the short storage case, and $\gamma_{2:2}^{D} = 0.45 \pm 0.02$ for the long storage case.   In each set, the $\approx$~$5\%$ difference between the short and long values is consistent with the extra fiber attenuation associated with the longer storage line, combined with small differences in patchcord connector loss. Importantly, we note that the $\gamma_{2:2}^{H}$ and $\gamma_{2:2}^{D}$ values are equal to within experimental uncertainty for both the short and long fiber-loop cases. This implies that the two basis states in an arbitrary qubit $\ket{\psi} = \alpha \ket{H} + \beta \ket{V}$ are equally attenuated during storage, resulting once again in an overall loss without a significant degradation in polarization qubit fidelity. 

Consequently, using Eq.~\ref{eq:eff}, we roughly summarize the end-to-end memory efficiency scaling for all polarization states as $\eta_{N\geq1}^{S} = 0.419 \times 0.50^{N-1} \times 0.662 \approx 0.55 \times 0.50^{N}$ for the short storage case, and $\eta_{N\geq1}^{L} = 0.398 \times 0.44^{N-1} \times 0.662 \approx 0.60 \times 0.44^{N}$ for the long storage case. Because the pass-through efficiency $\eta_0$ (i.e. switch off; no storage line, optical path $\approx$ 10.7~ns) happens to be $\gamma_{1:3} = 0.541 \pm 0.002$, we can further simplify the overall efficiency scaling estimate for the short storage case as $\eta_{N}^{S}\sim\!0.5^{N+1}$ for all $N$.

\begin{figure*}[htb!]
    \centering
    \includegraphics[width=0.75\textwidth]{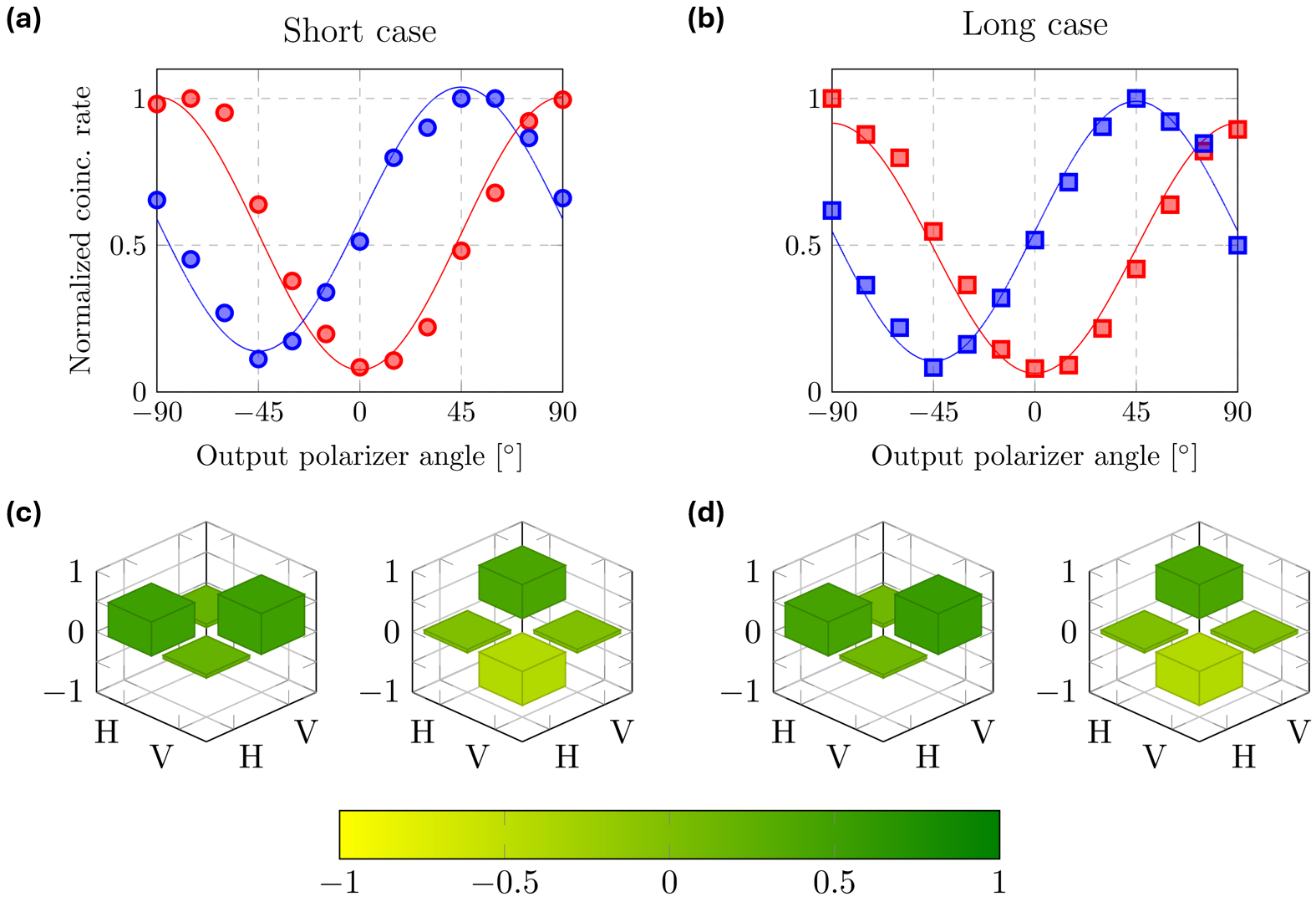}
    \caption{Example results from output qubit polarization characterization measurements for $N=3$, corresponding to 109.5~ns and 1.58$~\mu$s of total storage time for the short and long cases respectively. (a) and (b): Normalized Malus' law coincidence counting data for input states of $\ket{H}$ (red) and $\ket{D}$ (blue) for short and long storage line cases, respectively. (c) and (d): Reconstructed single photon density matrices for an input state of $\ket{R}$ for the short and long storage cases, respectively. In each pair of plots, the left and right matrices represents the real and imaginary parts of $\ket{R}$ respectively.} 
    \label{fig:pgfPlot:N=3subsetData}
\end{figure*}

For all three input single-photon polarization states $ \{\ket{H}, \ket{D},\ket{R} \}$ stored for $N$ cycles, we characterize the fidelity of the output state by experimentally reconstructing the single qubit density matrices, retrieved from $D_{1}$:$D_{2}$ coincidence counting measurements by maximum likelihood estimation~\cite{measofQubits}. In addition, for the case of the two linearly polarized input states, we use the visibility of fits to Malus' law coincidence counting data as a supplemental measure of output state quality. Figure~\ref{fig:pgfPlot:N=3subsetData} shows typical example results from both types of measurements for the case of $N=3$.  Panels (a) and (b) show output Malus' law curves for input states $\ket{H}$ (red) and $\ket{D}$ (blue) for the short and long storage cases, respectively, with fits to the data giving visibilities of $V_{H}^{S}=(86.09 \pm 1.59)\%$, $V_{D}^{S}=(76.46 \pm 2.12)\%$, $V_{H}^{L}=(86.99 \pm 1.92)\%$, and $V_{D}^{L}=(80.78 \pm 2.43)\%$. Panels (c) and (d) show the reconstructed density matrices for an input state of $\ket{R}$ for the short and long storage case, giving $F_{R}^{S}=(90.0 \pm 1.5)\%$, and $F_{R}^{L}=(89.4 \pm 1.9)\%$.

\begin{figure*}[hbt!]
    \centering
    \includegraphics[width=0.7\textwidth]{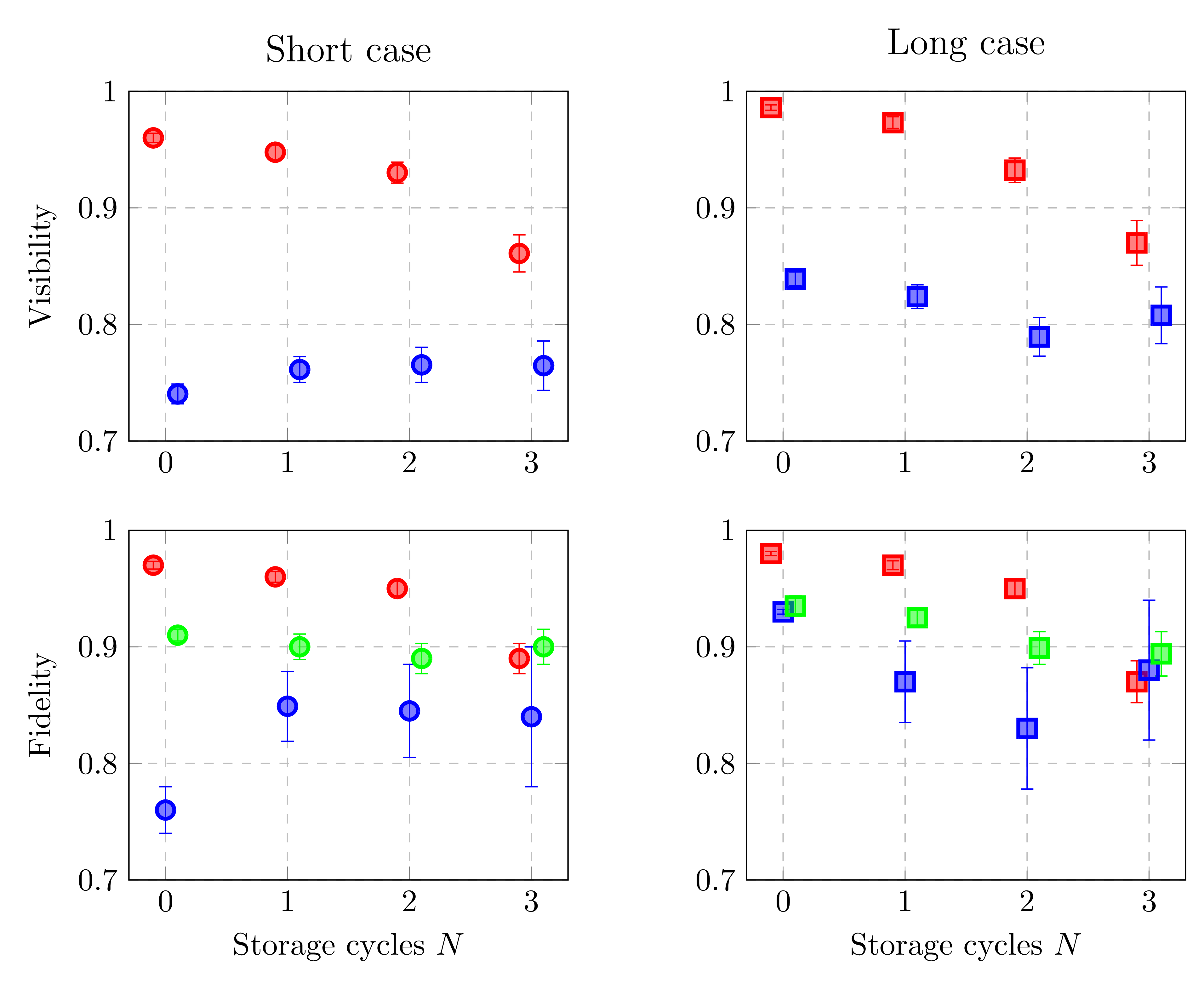}
    \caption{Summary of output state quality vs. memory storage time for both the short and long storage cases. (a) and (b): Best-fit Malus' law visibilities vs. $N$ for $\ket{H}$ (red) and $\ket{D}$ (blue) input states. (c) and (d): Fidelities vs. $N$ for $\ket{H}$ (red), $\ket{D}$ (blue), and $\ket{R}$ (green) input states. The data demonstrates very little polarization state degradation with increasing storage time. }
    \label{fig:pgfPlot:visANDfidScatterplot}
\end{figure*}

Figure~\ref{fig:pgfPlot:visANDfidScatterplot} summarizes the results of analogous visibility and fidelity measurements for these states for $N\in[0,3]$.  The error bars on the visibilities are retrieved from the fits to the data, while those on the fidelity data points represent standard deviations retrieved by using the four coincidence counting values for each tomographic measurement as mean values of Poissonian distributions, and then running Monte Carlo simulations with $\sim \!$$10^4$ samples. The larger uncertainties with increasing $N$ are simply due to a memory loss-induced decrease in coincidence counts for the same acquisition time (60~seconds per datapoint). 

It is clear from Figure~\ref{fig:pgfPlot:visANDfidScatterplot} that this fiber-coupled LAS platform causes very little polarization state degradation with increasing storage time. Note that in order to simulate realistic quantum memory network operation where the input state is unknown, the entire dataset comprising all three input states was collected without any adjustments to the memory alignment. In addition, when `plug-and-play' switching between the short and long storage case data sets, only the FPC in the reconfigurable storage line was adjusted. In this sense, the $\ket{H}$, $\ket{D}$, and $\ket{R}$ results shown in Fig.~\ref{fig:pgfPlot:visANDfidScatterplot} demonstrate the inherent robustness against polarization errors in this hybrid memory platform.

\section{\label{sec:conclusion}Summary and outlook}
We have demonstrated a fully fiber-coupled LAS absorptive quantum memory device for ultrabroad-bandwidth polarization-encoded qubits at 780 nm. We achieved a memory efficiency that scales as $\sim$$0.5^{N+1}$, where $N$ is the number of storage cycles, and observed average qubit output fidelities of $\sim\!90\%$ that do not significantly degrade with $N$ due to a combination of intrinsic bit flip and phase shift error mitigation effects. In addition, an intentional bit flip operation applied within the storage line converted the overall setup into a large common-path figure eight (F8) interferometer that facilitated hours-long, drift-free data runs. 

In principle, the combination of a high-speed switch (here, $\sim$10 ns rise time, $\sim$1 MHz repetition rate) and the easily reconfigurable length of a fiber storage `loop' enables seamless LAS memory operation over timescales ranging from the ultrafast regime for high qubit accessibility to the ultralong storage regime for extended memory lifetimes.  We demonstrated this static reconfigurability architecture by observing high-fidelity qubit storage for up to $N=3$ storage cycles in examples of both regimes; one with high qubit accessibility (every $\sim$40 ns) and one with long overall storage time $(\sim$1.5 $\mu$s). 

Given the overall loss value of $\sim$50\% per cycle that limited $N$ in this proof-of-concept work, it is interesting to speculate on the kinds of improved loss budgets that could potentially be achieved with current technology. For example, our implementation suffered from $\approx$~10\% loss per pass through the PC, but this value can be easily driven down to $\sim\!$1\% using RTP-based PCs with carefully designed anti-reflection (AR) coatings~\cite{kaneda_quantum-memory-assisted_2017}. In addition, the $\approx$~19\% loss in our fiber retro-reflector and the $\approx$~15\% ejection loss due to imperfect polarization rotations in the free-space circulator could be realistically driven down to $\sim$2\% each, while the nominal $\sim$7.5\% loss per fiber connection could be virtually eliminated by proper fusion splicing.  Furthermore, the free-space to SMF coupling efficiencies (implemented here with manual three-axis micro-positioners) within the various measured $\gamma_{i:j}$ values had an average of $(87.0 \pm 1.8)\%$, and these values could be expected to surpass $95\%$ using AR-coated fibers and six-axis nano-positioning systems. 

The combination of all of these improved loss values in our setup would increase the memory efficiency from $\sim$$0.5^{N+1}$ up to $\sim$$0.9^{N+1}$ for our short storage case, and $\sim$$0.8^{N+1}$ for our long storage case. In addition, utilizing these same loss values but switching from 780~nm operation (fiber attenuation of $\sim$4~dB/km) to 1550~nm operation (fiber attenuation of $\sim$0.2~dB/km) would result in a similar memory efficiency scaling of $\sim$$0.9^{N+1}$ for ultrafast applications, but significantly enhanced performance for ultralong storage cases.  For example, using a 5~km delay line ($\Delta\tau_{l}$ $\sim$50~$\mu$s)  would give an efficiency scaling of only $\sim$$(10^{-4})^{N+1}$ at 780~nm, but $\sim$$0.5^{N+1}$ at 1550~nm. However, it is important to note that even the highest value quoted above ($\sim$$0.9^{N+1}$) corresponds to a $1/e$ memory lifetime of only a few cycles.  In order to obtain $1/e$ lifetimes on the order of, say, 100 cycles, efficiency scalings of $\sim$$0.99^{N+1}$ would be required. In principle, such high performance can be achieved with free-space delay lines using ultra-low loss mirrors~\cite{freeSpaceLoopQM_8_2022, sun2025freeSpaceLoop8} rather than fiber, albeit at the expense of losing the easy plug-and-play reconfigurability demonstrated here.

Finally, we note that the polarization-insensitive hybrid LAS quantum memory presented here can be used without any changes for the storage of arbitrary time-bin qubits, provided the `long vs. short' separation time (typically tens of ps~\cite{timebinsep_2022}) is sufficiently shorter than the minimum storage cycle time (here, $\sim$10 ns). In addition, the memory can be used without changes for storage of frequency-bin qubits, provided the `red vs. blue' detuning (typically tens of GHz~\cite{freq-bin-sep-2023}) is smaller than the intrinsic bandwidth of the switch (here, $>$10~THz). This remarkable flexibility in photonic qubit encoding, combined with the easily reconfigurable storage time concept demonstrated here, may render fiber-coupled LAS memories useful for a variety of near-term quantum networking applications. Prospects for multi-mode operation~\cite{MM-theory-2007} and dynamical reconfigurability~\cite{freeSpaceLoopQM_8_2022} may even further expand the application space of the overall LAS quantum memory platform.

\textbf{Acknowledgements:} This work was supported in part by National Science Foundation under Grant Award Number 2013464 and National Institute of Standards and Technology under Grant Number 60NANB24D106. S.C. thanks Adrian Udovi\^{c}i\'{c} for helpful discussions on quantum state tomography.

\bibliography{0main}
\end{document}